\documentclass[twocolumn,showpacs,preprintnumbers,amsmath,amssymb]{revtex4}

\usepackage[dvips]{hyperref}
\usepackage{graphicx}

\begin{document}

\title{High-pressure transport properties of CeRu$_2$Ge$_2$}

\author{H. Wilhelm$^1$,
D. Jaccard$^2$, V. Zlati\'c$^3$,
R. Monnier$^4$, B. Delley$^5$, and B. Coqblin$^6$}

\affiliation{$^1$Max--Planck--Institut f\"ur Chemische Physik fester Stoffe,
  N\"othnitzer Str. 40, 01187 Dresden, Germany}

\affiliation{$^2$D\'epartement de Physique de la Mati\`ere Condens\'ee,
   Universit\'e de Gen\`eve, Quai Ernest-Ansermet 24, 1211 Gen\`eve 4,
   Switzerland}

\affiliation{$^3$Institute of Physics, Bijeni\v{c}ka cesta 46, P. O. Box
304, 10001 Zagreb, Croatia}

\affiliation{$^4$ETH H\"onggerberg, Laboratorium f\"ur
Festk\"orperphysik,
8093 Z\"urich, Switzerland}

\affiliation{$^5$Paul Scherrer Institut, WHGA/123, 5232 Villigen PSI,
Switzerland}

\affiliation{$^6$Laboratoire de Physique des Solides, Universit\'e
Paris-Sud, B\^at. 510, 91404 Orsay, France}

\begin{abstract}
The pressure-induced changes in the temperature-dependent thermopower
$S(T)$ and electrical resistivity $\rho(T)$ of CeRu$_2$Ge$_2$ are
described within the single-site Anderson model. The Ce-ions are
treated as impurities and the coherent scattering on different
Ce-sites is neglected. Changing the hybridisation $\Gamma$ between the
4f-states and the conduction band accounts for the pressure
effect. The transport coefficients are calculated in the non-crossing
approximation above the phase boundary line. The theoretical $S(T)$
and $\rho(T)$ curves show many features of the experimental data.  The
seemingly complicated temperature dependence of $S(T)$ and $\rho(T)$,
and their evolution as a function of pressure, is related to the
crossovers between various fixed points of the model.
\end{abstract}

\maketitle

\section{Introduction}
The thermoelectric power ($S$) of Ce-based heavy fermion (HF) or
unstable valence compounds and alloys exhibits a seemingly complicated
temperature dependence, $S(T)$. Depending on the hybridisation
($\Gamma$) between the 4f and the conduction electrons, the systems can
attain ground states varying from magnetically ordered to mixed valent
(MV), like CeCu$_2$Ge$_2$ \cite{Jaccard92} and CeNi$_2$Si$_2$
\cite{Levin81}, respectively. Between these two extreme cases magnetic
Kondo systems, e.g.~CeAl$_2$ \cite{Jaccard82}, and HF compounds like
CeCu$_2$Si$_2$ \cite{Jaccard85}, CeRu$_2$Si$_2$ \cite{Amato89}, and
CeCu$_6$ \cite{Amato87} are situated. A systematic development of
pronounced features in $S(T)$ becomes apparent if the compounds are
arranged according to increasing $\Gamma$-values, i.e. upon
approaching the MV regime.

One possibility to increase $\Gamma$ in Ce-compounds is to apply
pressure ($p$).  Depending on the compounds' ground-state properties
(i.e.~the value of $\Gamma$) at ambient pressure, some of the
anomalies at low and high temperature evolve and their behaviour under
pressure can be studied. Examples for the influence of pressure on
$S(T)$ of a magnetically ordered compound are CeCu$_2$Ge$_2$ and
CePd$_2$Si$_2$ \cite{Jaccard92,Link96b,Link96a}. In the case of HF
systems, pressure effects on $S(T)$ of CeAl$_3$ and CeCu$_2$Si$_2$
\cite{Fierz88,Jaccard85} were explored intensively. The transport
properties of the magnetically ordered CeRu$_2$Ge$_2$ clearly revealed the
pressure-induced development of two well resolved maxima in $S(T)$
over a considerably large pressure range \cite{Wilhelm04}. The
development of a pronounced positive maximum close to room temperature
is related to the splitting of the 4f states due to the crystalline
electric field (CEF) \cite{bhattacharjee76}. The non-monotonic $S(T)$
below about 10~K is very likely caused by the occurrence of magnetic
order and the opening of a gap in the magnetic excitation
spectrum. Upon approaching the magnetic/non-magnetic phase boundary at
a critical pressure $p_c=7.8$~GPa, a low-temperature maximum evolves
above 10~K. It is very likely related to the Kondo effect and gives a
measure of the Kondo temperature, $T_K$. Well above this pressure only
the high-temperature maximum remains and the $S(T)$-curves lose their
complexity.

The anomalous $S(T)$-data of Ce- and Yb-based compounds were subject
of many theoretical investigations and are still discussed
controversially (see Ref.~\cite{Zlatic03} and references therein). In
the present article we present a qualitative description of the
pressure-induced changes in $\rho(T)$ and $S(T)$ of CeRu$_2$Ge$_2$
within the single-site Anderson model. The spectral function of the
4f-electron was obtained within the non-crossing approximation
(NCA). The results above the characteristic energy scale $T_0$ can be
used reliably, whereas for $T<T_0$ the limitations of the method
should be kept in mind. An accurate solution of the single-impurity
model at low temperature is not useful, because we are dealing with a
stoichiometric compound, which is magnetically ordered at low
pressure. A single-site model cannot describe ordered compounds below
the coherence temperature anyway. Despite these limitations, the
Anderson model yields a qualitative description of the experimental
data of CeRu$_2$Ge$_2$ above the ordering temperature.

\section{Experimental results}
\label{sec:results}
A clamped Bridgman anvil cell with synthetic diamonds was used to
pressurise the sample up to 16~GPa. The four-point $\rho(T)$ and the
$S(T)$ measurements were carried out on {\it one} sample in the
temperature range $1.2~{\rm K}<T<300~{\rm K}$. It is important to
note, that $\rho(T)$ as well as $S(T)$ were measured perpendicular to
the tetragonal $c$-axis. A general description of the experimental
high-pressure set-up for $S(T)$ measurements can be found in
Ref.~\cite{Jaccard98}. The experimental results of interest in the
present context are summarised in the following. A thorough
presentation of the experimental findings can be found in
Ref.~\cite{Wilhelm04}.

\begin{figure}
\begin{center}
\includegraphics[width=1\columnwidth,clip]{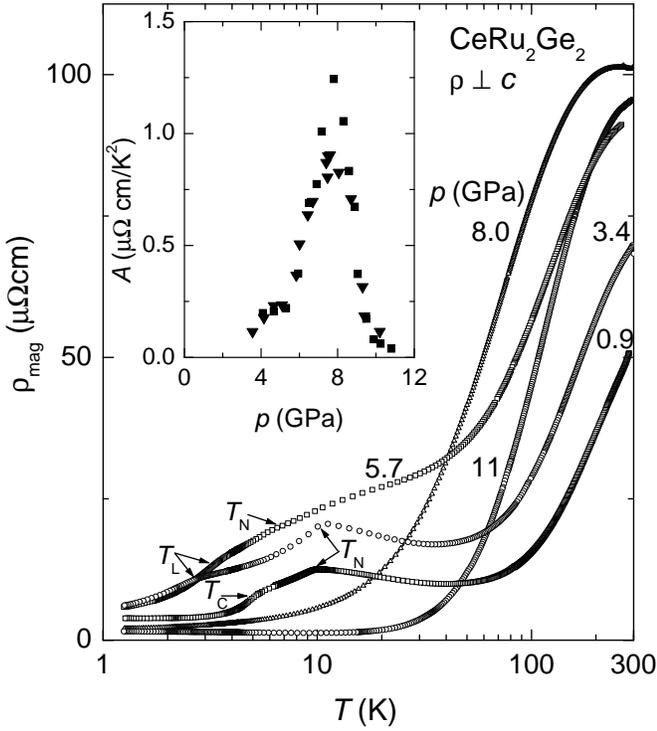}
\caption{Magnetic contribution $\rho_{mag}(T)$ to the electrical
   resistivity of CeRu$_2$Ge$_2$ at different pressures. Two different
   antiferromagnetic phases occur below $T_N$ and $T_L$, and a
   ferromagnetic ground state is present below $T_C$ and low
   pressure. No traces of magnetic order are observed for $p>7$~GPa
   above 1.2~K. Inset: Pressure dependence of the $A$-coefficient
   obtained from a fit of $\rho(T)=\rho_0 + AT^2$ for $T< 0.5$~K to
   the data of Ref.~\cite{Wilhelm99}.}
\label{fig:rhomag}
\end{center}
\end{figure}

\begin{figure}
\center{\includegraphics[width=1\columnwidth,clip]{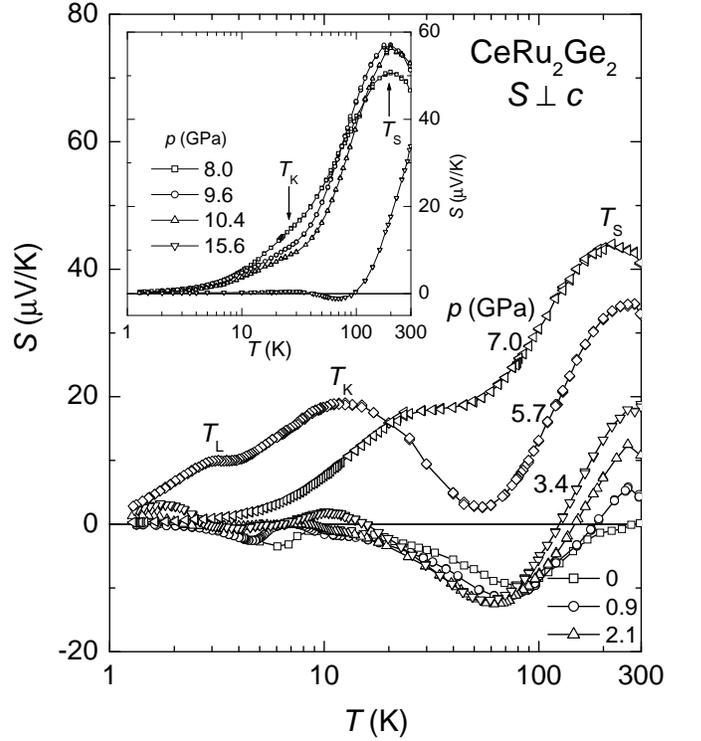}}
\caption{Temperature dependence of the thermoelectric power $S(T)$ of
   CeRu$_2$Ge$_2$ for various pressures. The features present in the low
   temperature part of $S(T)$ for $p\leq 5.7$~GPa are due to the
   magnetic order (e.g. $T_L$). $T_K$ and $T_S$ label the centre of
   broad, pressure-induced maxima, related to the Kondo effect and the
   crystalline electric field, respectively. The inset shows $S(T)$
   data of CeRu$_2$Ge$_2$ in the non-magnetic phase.}
\label{fig:tepall}
\end{figure}

The magnetic part $\rho_{mag}(T)$ of the total electrical resistivity,
$\rho(T)$, is shown in Fig.~\ref{fig:rhomag}. It was obtained by
subtracting an appropriate phonon contribution \cite{Wilhelm04}. In
the region of interest here $(T>10~K)$, a maximum in $\rho_{mag}(T)$
develops near room temperature for intermediate pressures. Its origin is
attributed to the Kondo exchange interaction between the conduction
electrons and the crystal-field split ground state of the
Ce$^{3+}$-ions.

This interpretation is supported by the evolution of a high-temperature
maximum in $S(T)$ as can be seen in Fig.~\ref{fig:tepall}. Its position,
$T_S$, corresponds to a fraction of the crystal-field splitting (the first
excited doublet is at $\Delta=500$~K \cite{Felten87,Loidl92}) and its
amplitude increases linearly with pressure up to a value of $55~\mu{\rm V/K}$
at about 10~GPa. As in $\rho(T)$ the features in $S(T)$ below about 10~K and
$p\leq 3.4$~GPa reflect presumably the appearance of long-range magnetic order
\cite{Wilhelm04}. At 5.7~GPa, however, a pronounced maximum at $T_K\approx
12$~K is present. Its position is very sensitive to pressure and a trace of it
can be anticipated at about 40~K (see 8.0~GPa data shown in the inset to
Fig.~\ref{fig:tepall}).

\begin{figure}
\center{\includegraphics[width=1\columnwidth,clip]{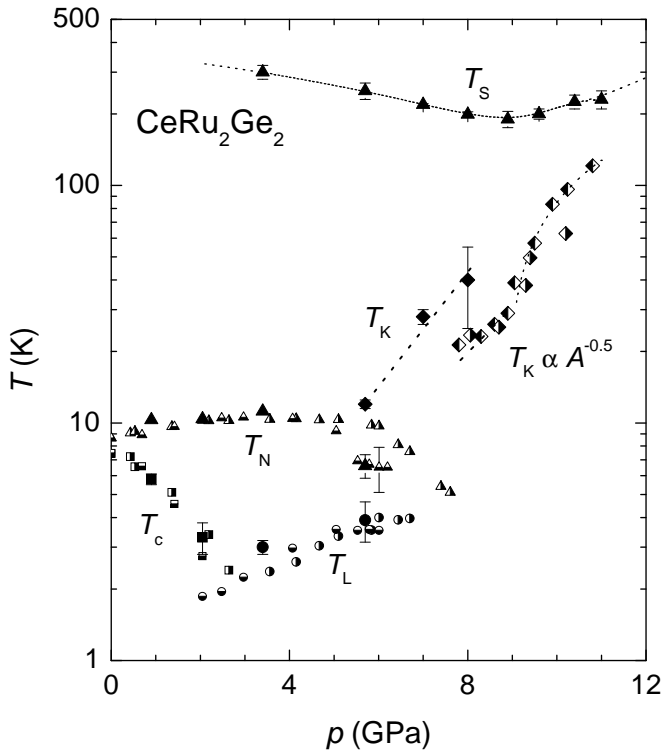}}
\caption{Detailed ($T,p$) phase diagram of CeRu$_2$Ge$_2$ as obtained from
   electrical resistivity \cite{Wilhelm99,Wilhelm98}, calorimetric
   \cite{Bouquet2000}, and the combined $\rho(T)$ and $S(T)$ (bold
   symbols) measurements \cite{Wilhelm04}. The long-range magnetic
   order ($T_N$ and $T_L$: antiferromagnetic, $T_C$: ferromagnetic) is
   suppressed at $p_c \approx 7.8$~GPa. $T_S$ and $T_K$ represent the
   centre of peaks in $S(T)$.}
\label{fig:phasediagram}
\end{figure}

Based on these results as well as earlier transport
\cite{Wilhelm99,Wilhelm98} and calorimetric measurements
\cite{Bouquet2000} a $(T,p)$ phase diagram of CeRu$_2$Ge$_2$ can be
drawn (Fig.~\ref{fig:phasediagram}). In the present context, the main
observation is that the long-range magnetic order is suppressed at a
critical pressure $p_c\approx 7.8$~GPa and that a HF like behaviour
equivalent to that in CeRu$_2$Si$_2$ at ambient pressure sets in for
$p>p_c$.  The almost identical $S(T)$ curves of CeRu$_2$Si$_2$ at
ambient pressure and CeRu$_2$Ge$_2$ at high pressure strongly suggest
that the unit-cell volume is the crucial parameter as far as the size
of the Kondo exchange interaction is concerned, and suggests that the
position of the peak in $S(T)$ at low temperature is a measure for the
Kondo temperature $T_K$ \cite{Wilhelm04}. Additional support for this
assignment is provided by the pressure dependence of the
$A(p)$-coefficient (inset to Fig.~\ref{fig:rhomag}) of the
quadratic-in temperature $\rho(T)$ behaviour fitted to the data below
0.5~K reported in Ref.~\cite{Wilhelm99}. In the non-magnetic regime
the assumption $T_K\propto 1/\sqrt{A}$ holds and the calculated
$T_K(p)$ data are shown in Fig.~\ref{fig:phasediagram}
\cite{Wilhelm04}. They were normalised in such a way that at
$p=7.8+0.6$~GPa the value of $T_K=24$~K of CeRu$_2$Si$_2$ at ambient
pressure was obtained \cite{Besnus85}. The small pressure offset
corresponds to the fact that long-range magnetic order in
CeRu$_2$(Si$_{1-x}$Ge$_x$)$_2$ vanishes at a critical concentration
$x_c\approx 0.06$ \cite{Wilhelm04,Haen2000}.

\section{Theoretical model}

The theoretical description is based on an effective impurity
model. It is assumed that the scattering of conduction electrons on a
given Ce-ion depends on other Ce-ions only through the modification of
the conduction band. Fluctuation between 4f$^0$ and 4f$^1$
configurations by exchanging electrons with the conduction band are
allowed. The energy difference between the two configurations is
$|E_f|$ and the hopping is characterised by the matrix element
$V$. The 4f$^1$ configuration is represented by the CEF states which
are split by an energy $\Delta\ll |E_f|$. The local symmetry is taken
into account by specifying the degeneracy of the CEF levels. The
4f$^2$ configuration is excluded, i.e., an infinite strong repulsion
between 4f-electrons is assumed.

\subsection{Model hamiltonian}
\label{sec:hamiltonian}
Taking the impurity concentration equal to one, and
imposing that the average number of conduction
electrons and 4f-electrons is conserved (self-consistency condition)
\cite{Zlatic03}, the Hamiltonian for the
effective single-ion Anderson model is given by

\begin{equation}
H_{A}=H_{band}+H_{imp}+H_{mix}~,
\end{equation}

\noindent where $H_{band}$ describes the conduction states, $H_{imp}$
represents CEF states at $E_f$ and $E_f^*=E_f+\Delta$, and $H_{mix}$
gives the transfer of electrons between 4f and conduction states. All
energies are measured with respect to the chemical potential
$\mu$. The properties of the model depend in an essential way on the
CEF splitting $\Delta$ and the coupling constant $g=\Gamma/(\pi
|E_f|)$, where $\Gamma=\pi V^2/W$ measures the coupling strength
between the 4f-electrons and a semi-elliptical conduction band
(centred at $E_0>0$) of half-width $W$. In the limit $\Gamma \ll 1$
and $\Gamma <\Delta$, the model represents the 4f$^1$ configuration of
the Ce ion at ambient pressure and high temperature.  An increase of
pressure stabilises the 4f$^0$ configuration of the Ce-ion and
enhances the configurational mixing. Therefore, the influence of
pressure is accounted for by an increase of the hybridisation from
$\Gamma <\Delta$ to $\Gamma > \Delta$. In a Kondo-lattice compound the
hybridisation can give rise to charge fluctuations and in order to
obtain charge neutrality, $\mu$ was adjusted at each temperature and
hybridisation. Furthermore, it is assumed that pressure does not
change $\Delta$ and $W$ but shifts the bare f-level and the centre of
the conduction band by the same amount. Thus, for a given Ce-compound,
$n_c + n_f$ and $|E_0-E_f|$ remain constant at all temperatures and
pressures \cite{Zlatic05}.

In order to describe the influence of pressure on $S(T)$ and $\rho(T)$
of CeRu$_{2}$Ge$_2$, which is paramagnetic at high temperature and
ambient pressure, it is sufficient to consider the case $g\ll 1$. The
behaviour of the Anderson model in that limit is controlled by
several well defined fixed points \cite{Hewson93}. In the case of
$\Gamma < \Delta$, the impurity appears to be magnetic and the
low-temperature behaviour is characterised by the FL fixed point,
which describes a singlet formed by the 4f-moment and the conduction
electrons. An increase of temperature gives rise to a transition to
the local moment (LM) fixed point, which characterises a CEF split
4f-state weakly coupled to the conduction band. This transition takes
place around the Kondo temperature, $T_0$, determined by $g$,
$\Delta$, and the degeneracy of the CEF states
\cite{BCW87,Bickers87}. The definition of $T_0$ in the NCA is given in
the next section. The remarkable feature of the Anderson model is that
the physics at the FL and the LM fixed points is governed by the same
characteristic energy scale $T_0$. A further transition to a fixed
point related to the scattering of conduction electrons on an
effective six-fold degenerate 4f-multiplet occurs at temperatures
above $\Delta$. In the case of $\Gamma \gg \Delta$, the impurity is in
the non-magnetic MV regime.

The electrical resistivity and the thermopower of the single-impurity
Anderson model are obtained from the usual expressions \cite{Mahan81},
\begin{equation}
{\rho_{mag}}=\frac{1}{e^2 L_{0}}~,
\label{eq:conductivity}
\end{equation}

\begin{equation}
S=-\frac{k_B}{|e|T}\frac{L_{1}}{L_{0}}~,
\label{eq:thermopower}
\end{equation}

\noindent with $k_B$ the Boltzmann constant and $e$ the electronic
charge. The transport coefficients $L_{0}$ and $ L_{1}$ are given by
the static limits of the current-current and current-heat current
correlation function, respectively. The vertex corrections vanish and
the transport integrals can be written as \cite{BCW87,Bickers87,Mahan98},

\begin{equation}
L_{n}=\frac{\sigma_0}{e^2}\int_{-\infty}^{\infty}d\omega \left (
-\frac{df(\omega)}{d\omega} \right ) \tau(\omega)\omega^n~,
\label{eq:lij_final}
\end{equation}

\noindent where $\tau(\omega)$ is the conduction electron scattering rate
at energy $\omega$ \cite{BCW87,Bickers87},

\begin{equation}
\frac 1 {\tau(\omega)}= c N \pi V^2 A(\omega),
\label{eq:tau}
\end{equation}

\noindent with $A(\omega)=\mp \frac{1}{\pi}\mbox{Im}\; G_f(\omega \pm
i 0^{+})$ the 4f-electron spectral function, $G_f(\omega)$ the
retarded Green's function, $f(\omega)=1/[1+\exp(\omega/(k_BT))]$ the
Fermi function, $N$ the total number of scattering channels,
$\sigma_0$ a material-specific constant, and $c=1$ the concentration
of 4f-ions. $G_f(\omega \pm i 0^{+})$ is obtained by the NCA,
following closely Refs.~\cite{BCW87,Bickers87,Monnier90}.

The NCA results for CeRu$_2$Ge$_2$ are obtained for a ground state
doublet and an excited quartet split by $\Delta=0.07$~eV. It is
convenient to assume an excited quartet CEF state instead of two
excited doublets CEF levels. A significant change of the NCA results
does not occur, since the separation between the first and the second
excited state (0.021~eV) is less than their separation from the ground
state (0.043~eV) \cite{Felten87,Loidl92}. We used $E_f=-0.7$~eV,
$E_0=0.7$~eV, $W=4$~eV, and $\Gamma = 0.06$~eV to describe
CeRu$_{2}$Ge$_2$ at ambient pressure and high temperatures ($k_BT =
\Delta$), where the f-state is almost decoupled from the conduction
band and the renormalisation of the bare parameters is small. The
calculation yields the total number of 4f- and conduction electrons,
$n_{tot}=5.6301$, which is preserved at all subsequent calculations at
different temperatures and $\Gamma$-values by adjusting $\mu$
\cite{Zlatic05}.

\subsection{Spectral functions}
\label{sec:spectralfunction}

\begin{figure}
\center{\includegraphics[width=1\columnwidth,clip]{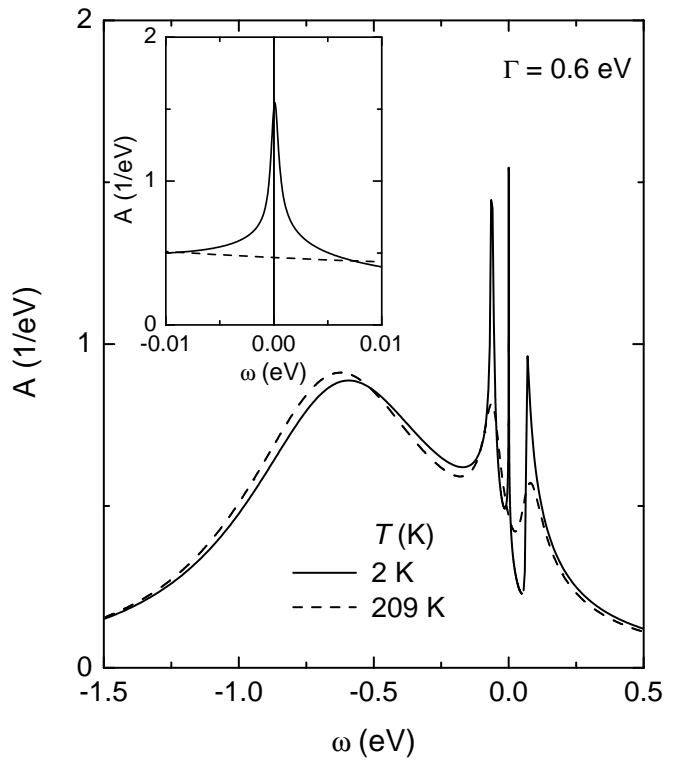}}
\caption{Spectral function $A(\omega)$, calculated for two
temperatures using $\Gamma=0.06$~eV and $\Delta=0.07$~eV. The broad
feature just above $E_f=-0.7$~eV is the charge-excitation peak. The
renormalised CEF peaks and the Kondo resonance appear close to the
chemical potential $\mu$ at $\omega=0$. The inset shows $A(\omega)$ in
the vicinity of $\omega=0$. In the low-temperature $A(\omega)$ the
Kondo peak is centred at $k_BT=\omega_0 \approx 0.2$~meV.}
\label{fig:zspectr_60}
\end{figure}

The spectral function $A(\omega)$ for a given value of $\Gamma$
(0.06~eV$<\Gamma < 0.2$~eV) is calculated for a discrete set of
temperatures in the range 2~K~$<T<800$~K. The change of the shape of
$A(\omega)$ with temperature for increasing $\Gamma$-values, i.e., for
enhancing the charge fluctuations, is presented in the
following. Although the precise relationship between $\Gamma$ and
pressure is not known it seems reasonable to assume that an increase
of $\Gamma$ results in an increase of pressure.

\begin{figure}
\center{\includegraphics[width=1\columnwidth,clip]{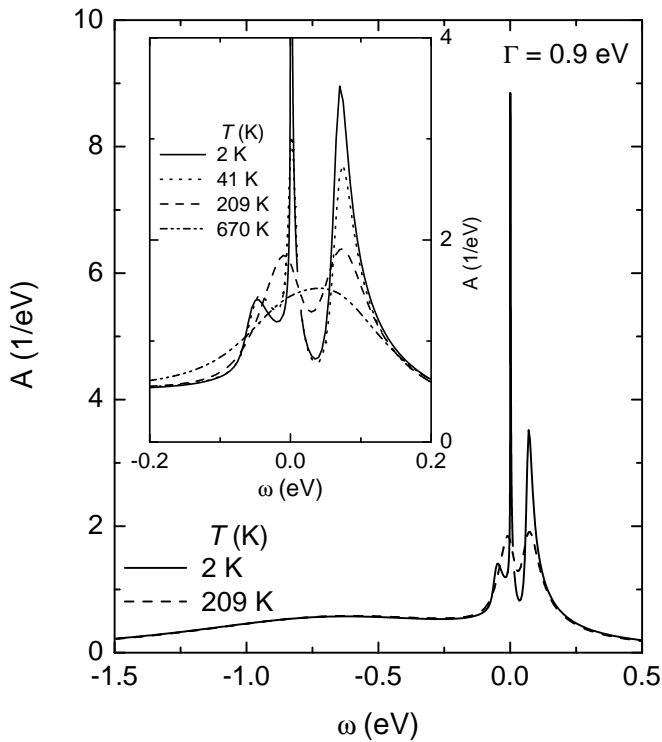}}
\caption{Spectral function $A(\omega)$ calculated with
$\Gamma=0.09$~eV and $\Delta=0.07$~eV for different temperatures. Here
the amplitude of the charge excitation and the lower CEF peak at
$\omega\approx \omega_0-\Delta$ are strongly reduced. The two features
above $\omega = 0 $ are the Kondo resonance and the renormalised CEF
peak at $\omega_0$ and $\omega_0 + \Delta$, respectively. The inset
shows the temperature variation of $A(\omega)$ in a smaller energy
range around the chemical potential.} 
\label{fig:zspectr_90}
\end{figure}

Figure~\ref{fig:zspectr_60} shows $A(\omega)$ at different
temperatures and for the case $\Gamma < \Delta$ and $n_f\simeq
1$. Common features to all curves are a broad charge-excitation peak
somewhat above $E_f$ (at $\omega\approx -0.59$~eV) and two features
due to spin excitations of the full CEF multiplet. For $k_BT < \Delta$
these two peaks grow and sharpen. At very low temperature, an
additional peak develops just above $\mu$ (inset to
Fig.~\ref{fig:zspectr_60}) and the low-energy part of $A(\omega)$ is
characterised by three pronounced peaks \cite{BCW87,Bickers87}. The
two peaks centred at about $\omega_0\pm\Delta$ are the renormalised
CEF peaks, while the peak at $\omega_0$, which is close to $\mu$, is
the Kondo resonance. It determines the low-temperature transport
properties. The energy $\omega_0$ provides the NCA definition of the
characteristic Kondo temperature, $T_0=\omega_0/k_B$. The comparison
with the numerical renormalisation group (NRG) calculations
shows~\cite{Costi96} that this $T_0$ is a reliable estimate of the
Kondo temperature even for a doubly degenerate Anderson model. Thus,
we assume that the NCA definition of $T_0$ provides the correct Kondo
scale of the CEF split single-ion Anderson model as well.

For $\Gamma \geq \Delta$ (and $0.9\leq n_f<0.95$) the relative
amplitudes of the peaks have changed (Fig.~\ref{fig:zspectr_90}). At
high temperatures, i.e. for $k_BT\simeq \Delta$, the only prominent
feature is the low-energy resonance which is due to the exchange
scattering on the full CEF multiplet ($A(\omega)$-curve at 670~K in
the inset to Fig.~\ref{fig:zspectr_90}). With decreasing temperature
the amplitude of the renormalised CEF peak above the chemical
potential and of the Kondo resonance increase. The larger $\Gamma$
with respect to Fig.~\ref{fig:zspectr_60} results in a reduced
amplitude of the charge excitation peak, in a broadening of the peaks,
and a shift of their positions to higher energies, whereas their
separation is still $\Delta$.

\begin{figure}
\center{\includegraphics[width=1\columnwidth,clip]{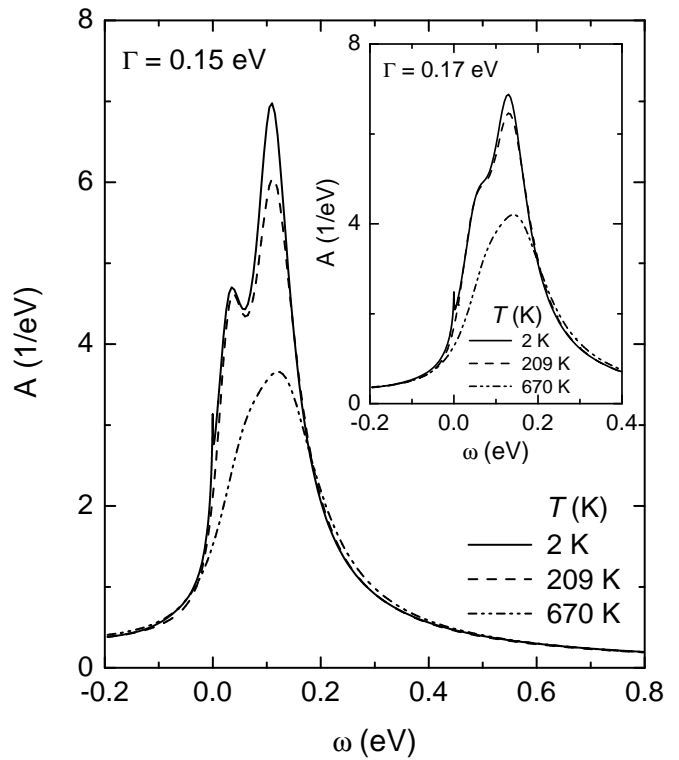}}
\caption{Spectral function $A(\omega)$ calculated for the parameters
$\Gamma=0.15$~eV and $\Delta=0.07$~eV at different temperatures. The
Kondo resonance and the CEF peak are separated for $T\ll 670$~K. The
data calculated for $T=2$~K exhibit an artifical spike at $\omega=0$
(see text). Inset: $A(\omega)$ calculated for $\Gamma=0.17$~eV and
$\Delta = 0.07$~eV. The Kondo peak occurs as shoulder at the
low-energy side of a broad peak centred at $\tilde{E}_f \approx 0.13$~eV.}
\label{fig:zspectr_150_170}
\end{figure}

A further increase of $\Gamma$, such that $0.75\leq n_f < 0.9$, shifts
the Kondo and the CEF peak to higher energies
(Fig.~\ref{fig:zspectr_150_170}). The Kondo peak is reduced but can
still be resolved as a small feature on the low-energy side of the
renormalised CEF peak. A different behaviour of $A(\omega)$ is found
for $\Gamma\gg 2\Delta$, i.e., $n_f < 0.8$ (inset to
Fig.~\ref{fig:zspectr_150_170}). It has only a single broad peak
centred at ${\tilde{E_f}} \approx 0.13$~eV, the renormalised position
of the virtual bound 4f-state, and $A(\omega)$ shows no splitting
despite the presence of the CEF-term in the Hamiltonian.  The shape of
this peak is retained at all temperatures and the Kondo resonance
appears only as a weak shoulder on the low-energy side of this broad
peak. Such a spectral function defines the MV regime of the Anderson
model, where the relevant energy scale is $k_B T_0={\tilde{E_f}}$,
which depends almost linearly on $\Gamma$ \cite{Zlatic05}. The
low-temperature part of $A(\omega)$ depicted in
Fig.~\ref{fig:zspectr_150_170} clearly shows the limitations of our
approach: The spike present at $\omega =0$ is an artefact of the NCA
and leads to an artifical enhancement of $\rho(T)$ and $S(T)$ at low
temperatures. The consequence of these non-analytic NCA states is
commented below.

\subsection{Calculated electrical resistivity and thermopower}
\begin{figure}
\center{\includegraphics[width=1\columnwidth,clip]{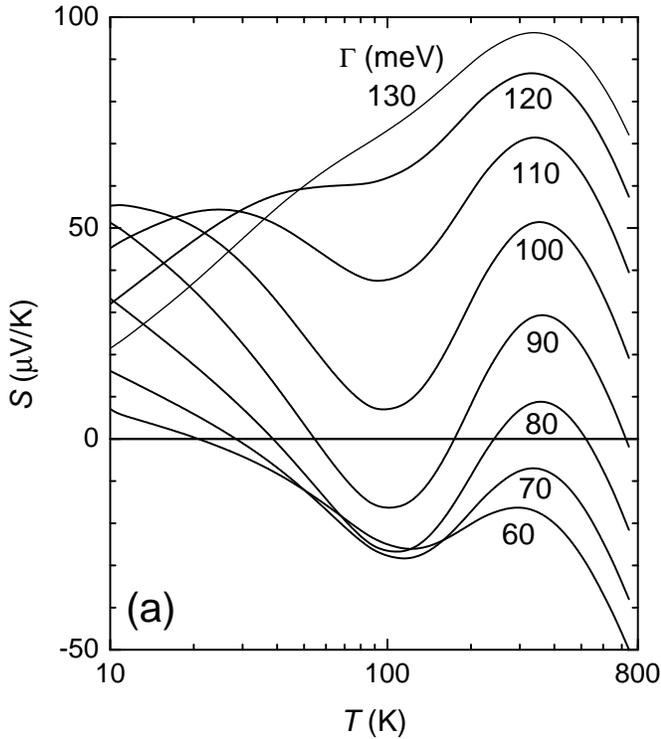}}
\center{\includegraphics[width=1\columnwidth,clip]{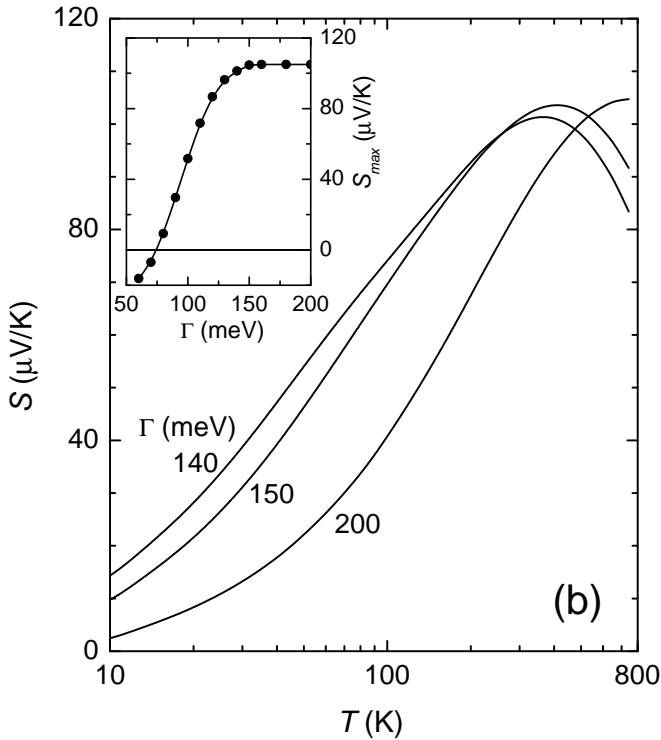}}
\caption{(a) Temperature dependence of the calculated thermopower
$S(T)$ for several small (a) and large (b) values of the hybridisation
$\Gamma$ and $\Delta = 0.07$~eV. The inset to (b) shows the
$\Gamma$-dependence of $S_{max}\equiv S(T_S)$, the amplitude of the
high-temperature maximum located at $T_S$ in the calculated
$S(T)$-data.}
\label{fig:theo_tep}
\end{figure}

The moments $L_n$ (with $n=0,1$) of the conduction electron scattering
rate, weighted with the energy derivative of the Fermi function
(eq.~(\ref{eq:lij_final})), determine the electrical resistivity
and the thermopower (eq.~(\ref{eq:conductivity}) and
(\ref{eq:thermopower})). Thus, $\rho(T)$ and $S(T)$ reflect the form
of $A(\omega)$ around the chemical potential within the Fermi window
$|\omega|<2k_B T$. The sign of $S(T)$ is positive if within the Fermi
window more states lie above than below the chemical potential. $S(T)$
is negative in the opposite case.

The NCA results for $S(T)$ and small $\Gamma$-values ($\Gamma \leq
100$~meV) show one well resolved maximum at $T_S\approx 300$~K and an
evolution of a second maximum, $T_{max}^{low}$, at low temperatures
(Fig.~\ref{fig:theo_tep}(a)). For 100~meV~$\leq \Gamma \leq 130$~meV,
the overall absolute value of $S(T)$ has increased and $T_S$ is
slightly shifted to lower temperatures. As a consequence of the
appearance of the low-temperature maximum, the minimum in $S(T)$
around 100~K becomes less pronounced. $T_{max}^{low}$ correlates with
the Kondo scale $T_0$. At high $\Gamma$-values, i.e.,
$\Gamma>130$~meV, the shape of $S(T)$ is characterised by the 
high-temperature maximum which is shifted upwards in temperature
(Fig.~\ref{fig:theo_tep}(b)). Its amplitude, $S_{max} \equiv
S(T_S)$, increases almost linearly with $\Gamma$ and
saturates above $\Gamma\approx 150$~meV (inset to
Fig.\ref{fig:theo_tep}(b)).

\begin{figure}
\center{\includegraphics[width=1\columnwidth,clip]{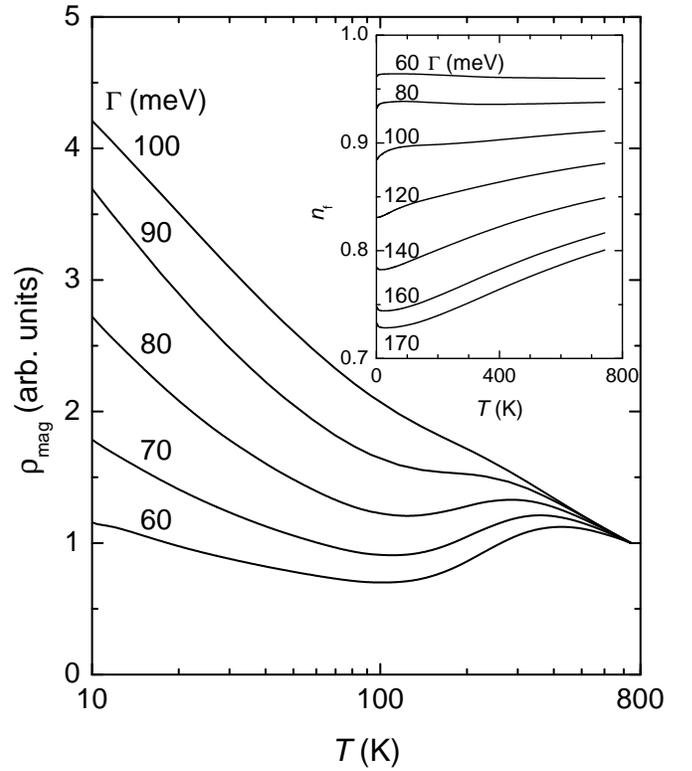}}
\caption{Calculated temperature dependence of the electrical
resistivity $\rho_{mag}(T)$ for different hybridisation $\Gamma$ and
$\Delta = 0.07$~eV. The enhancement of $\rho(T)$ at low temperature is
an artefact of the calculation (see text). The inset shows the
temperature variation of the f-electron number, $n_f$, at various
values of $\Gamma$.}
\label{fig:theo_rho}
\end{figure}

Figure~\ref{fig:theo_rho} depicts the calculated $\rho_{mag}(T)$ for
several (small) values of $\Gamma$. The low-temperature
$\rho_{mag}(T)$ exceeds that at high temperature, since the NCA
overestimates $A(\omega)$ at small $\omega$. Nevertheless, the
logarithmic increase, as temperature decreases towards $T_0$, is
preceded by the formation of a high-temperature maximum well above
$T_0$. Its position correlates very well with the maximum at $T_S$ in
$S(T)$. The maximum in $\rho_{mag}(T)$ becomes less pronounced and is
masked for $\Gamma>100$~meV due to the artifically large $A(\omega)$
around $\mu$. The occurrence of these non-analytic NCA states in
$A(\omega)$ for large $\Gamma$-values has a minor effect on the
4f-electron number, $n_f$ (inset to Fig.~\ref{fig:theo_rho}). $n_f$ is
the integral of $A(\omega)$ multiplied by the Fermi function. In the
MV regime ($\Gamma > 2 \Delta$) $n_f(T)$ decreases strongly from the
high-temperature limit to $n_f\approx 0.7$ at low temperatures. The
small increase at the lowest temperature results from the spike in
$A(\omega)$. In the Kondo regime ($\Gamma < \Delta$) $n_f\approx 1$ is
almost temperature independent. The variation of $n_f(T)$ with
$\Gamma$ reflects the change of regimes.

The strong increase of $\rho_{mag}(T)$ and the enhanced $S(T)$ at low
temperature (for $T\ll T_0$) is a consequence of the spike in
$A(\omega)$. This becomes particularly severe for large $\Gamma$
values ($\Gamma > 100$~meV). In this region the characteristic
temperature scale $T_0$ increases very rapidly with $\Gamma$ and
$A(\omega)$ acquires non-analytic NCA states already at rather high
temperatures (inset to Fig.~\ref{fig:zspectr_150_170}). The integral
$L_{0}$ is strongly underestimated, resulting in a $\rho_{mag}(T)$
which is too large (eq.~(\ref{eq:conductivity})). The integral $L_{1}$
is less affected, because the artificial NCA states are removed by the
additional $\omega$-factor in eq.~(\ref{eq:lij_final}). Thus, the
shape of $S(T)\simeq L_{1}^{NCA}/L_{0}^{NCA}$ seems to be
qualitatively correct, even for large $\Gamma$-values, but the
magnitude of the low-temperature $S(T)$ is enhanced, since
$1/L_{0}^{NCA} \gg 1/L_{0}^{exact}$. These difficulties are well known
\cite{BCW87,Bickers87} and can be solved easily in the Kondo limit,
where the model has a unique Kondo scale, $T_0$. The latter can be
calculated in the LM regime, where the NCA is reliable. Once $T_0$ is
known, the low-temperature transport properties can be inferred from
the universal power laws $\rho(T) \propto 1-a(T/T_0)^2$ and
$S(T)\propto b(T/T_0)$ (where $a$ and $b$ follow from the Sommerfeld
expansion) which hold in the FL regime. Thus, combining the NCA and
the FL theory, we can discuss the experimental data at all
temperatures at which the single-ion approximation holds.

\section{Discussion}
The evolution of the calculated transport coefficients described above
has several features of the pressure-induced changes in
$\rho_{mag}(T)$ and $S(T)$ of CeRu$_{2}$Ge$_2$ presented in
Sec.~\ref{sec:results}. In the following, the evolution of the
transport properties is related to the modifications of $A(\omega)$
caused by the increase of $\Gamma$ and compared to the experimental
data. However, the above mentioned limitations of the theoretical
approach should be kept in mind.

As far as the NCA results of $\rho_{mag}(T)$ are concerned, the
comparison of the theoretical and experimental data is limited to
temperatures above about 100~K. This is due to the fact that the NCA
yields an enhanced $A(\omega)$ at small $\omega$ and large $\Gamma$
and the calculated $\rho_{mag}(T)$ increase strongly towards low
temperatures. In addition, the single-site model leads to a
low-temperature saturation of $\rho_{mag}(T)$, which is not observed
in stoichiometric compounds. Thus, the NCA results for $\rho_{mag}(T)$
can only be used for a qualitative discussion of the experimental data
at low pressure. The clear high-temperature maximum and the subsequent
shallow minimum, present in $\rho_{mag}(T)$ for small hybridisation,
are seen in the experimental behaviour of $\rho_{mag}(T)$ for $p\leq
3.4$~GPa (Fig.~\ref{fig:rhomag}). Furthermore, the weak
$\Gamma$-dependence of the maximum position in the calculated
$\rho_{mag}(T)$ is also observed in the experimental data.

The clear features in the calculated $S(T)$-data and their evolution
with $\Gamma$ make a detailed comparison with the effect of pressure
on $S(T)$ possible. The ambient pressure $S(T)$ is negative for
temperatures below 300~K (Fig.~\ref{fig:tepall}). This shape of $S(T)$
corresponds to that calculated for $\Gamma = 60$~meV
(Fig.~\ref{fig:theo_tep}(a)). It results from the fact, that within
the Fermi window less spectral weight is present above than below
$\mu$. The difference, however, is small, which explains the small
negative value of $S(T)$.  For $k_BT < \Delta$ the amplitude of the
renormalised CEF peak in $A(\omega)$ below the chemical potential
increases (Fig.~\ref{fig:zspectr_60}) and $S(T)$ attains slightly
larger negative values in agreement with the ambient pressure data. At
even lower temperature, the development of the Kondo resonance leads
to more spectral weight above than below $\mu$ and $S(T)$ starts to
increase. This leads to the evolution of the minimum around 100~K and
$S(T)>0$ below 20~K. The minimum is clearly present in the
experimental data, whereas the occurrence of the long-range magnetic
order and the opening of a spin-gap \cite{Raymond99} conceals the
positive maximum at low temperature. The features in the experimental
$S(T)$ data below 10~K are related to magnetic order and cannot be
accounted for in the theory, since the single-ion approximation is not
suited for the ordered phase of stoichiometric compounds. Here the
4f-electrons become coherent at low enough temperatures, leading to
magnetic transitions or to the formation of a heavy FL, like in
CeRu$_{2}$Ge$_2$ at low or high pressure, respectively.

For $\Gamma > \Delta$, more spectral weight is present above than
below $\mu$.  In this case the exchange scattering on the full CEF
multiplet yields $S(T)> 0$ as experimentally observed for $p\geq
5.7$~GPa below 300~K. Reducing the temperature shifts spectral weight
below the chemical potential (Fig.~\ref{fig:zspectr_90}) and $S(T)$
decreases. This redistribution leads again to a small positive
low-temperature maximum at $T=T_{max}^{low}\simeq T_{K}^{H}$. The
minimum in the $S(T)$-curves can be slightly negative or positive,
depending on the value of $\Gamma$ with respect to $\Delta$. This
becomes evident for $\Gamma=90$~meV and 100~meV
(Fig.~\ref{fig:theo_tep}(a)) and in the experimental data for
$p=3.4$~GPa and $p=5.7$~GPa (Fig.~\ref{fig:tepall}). $S(T)$ exhibits
only a shallow minimum or a shoulder below $T_S$, like the
curve for $\Gamma=120$~meV in Fig.~\ref{fig:theo_tep}(a). The
equivalent experimental data are those measured at 7.0~GPa depicted in
Fig.~\ref{fig:tepall}.

\begin{figure}
\center{\includegraphics[width=1\columnwidth,clip]{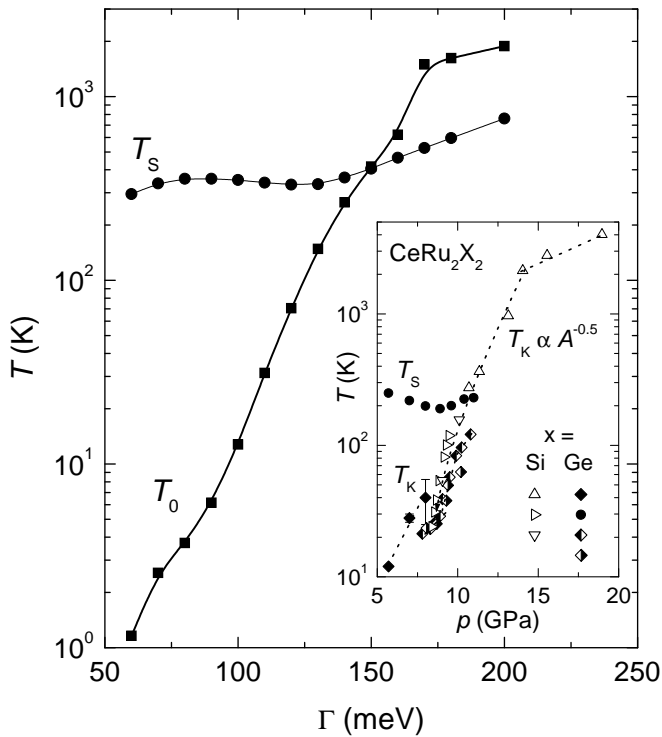}}
\caption{$T_S$, the position of the high-temperature
maximum in the calculated $S(T)$, and $T_0$, the characteristic
temperature scale, vs.  hybridisation $\Gamma$. $T_0$ corresponds to
$T_K$ (small $\Gamma$), $T_K^H$ (intermediate $\Gamma$), and
${\tilde{E_f}}$ (large $\Gamma$). The inset shows the experimental
data of $T_S$ and $T_K$ of CeRu$_2$Ge$_2$ and $T_K\propto \sqrt A$ of
CeRu$_2$Si$_2$ take from literature
\cite{Payer93,Mignot89,Thompson85}. The latter are shifted in pressure
by 8.4~GPa (see text).} \label{fig:theo_phasediagram}
\end{figure}

The Kondo resonance appears only as a shoulder at the broad
high-temperature maximum in the calculated $S(T)$-curves for $\Gamma >
2 \Delta$. In this region a drastic change of the relative spectral
weight of the peaks in $A(\omega)$ occurred and the Kondo resonance
represents only a negligible correction to the spectral weight within
the Fermi window. $S(T)$ is always positive and grows monotonically
towards the high-temperature maximum at $T_0\propto {\tilde{E_f}}$
(Fig.~\ref{fig:theo_tep}(b)) as experimentally observed for $p>10$~GPa
(inset to Fig.~\ref{fig:tepall}). It is noteworthy, that the almost
linear increase of the calculated value of $S(T_S)$ is in qualitative
agreement with the experimental findings (inset to
Fig.~\ref{fig:theo_tep}(b)) \cite{Wilhelm04}.

Figure~\ref{fig:theo_phasediagram} shows the $\Gamma$-dependence of
$T_0$ and $T_S$ determined from $A(\omega)$ and the calculated
$S(T)$-curves, respectively. The NCA values of $T_0$ obtained for
$\Gamma\ll \Delta$ are very well approximated by the scaling
expression $T_0= T_K(g)$ \cite{BCW87,Bickers87}.  For a ground state
doublet and an excited quartet it is $T_K(g) \simeq W g^{1/2}
\exp\{-1/(2g)\}(W/\Delta)^{4/2}$. The exponent 4/2 reflects the
assumption of an excited quartet instead of a doublet state. The Kondo
resonance is due to spin excitations of the CEF doublet, whereas the
Kondo temperature is enhanced by $(W/\Delta)^2$ due to the presence of
excited CEF states. For intermediate values of $\Gamma$
(e.g. $\Gamma\approx 100$~meV), the relationship between $T_0$ and the
coupling constant is $T_0=T_K^H \simeq W g^{1/2}
\exp\{-1/(2g)\}[W/(\Delta+T_K)]^{4/2}$.

The $T_0$-line in the $(T,\Gamma)$-phase diagram
(Fig.~\ref{fig:theo_phasediagram}) separates the FL region from the LM
region and the MV regime.  In the Kondo regime, an increase of
temperature at constant hybridisation gives rise to a transition from
the FL regime to the LM regime of a CEF ground state, and then to the
LM regime of a full CEF multiplet.  The crossover between the two LM
regimes is accompanied by the minimum of $S(T)$ (Fig.~\ref{fig:tepall}
and \ref{fig:theo_tep}).  At low coupling ($\Gamma \leq 90$~meV and
$T_0\simeq T_K$), the thermopower becomes negative at the crossover
from the FL to the LM regime and it stays negative in the LM regime of
a full CEF multiplet.  At intermediate coupling ($\Gamma \approx
100$~meV and $T_0\simeq T_K^H$), the FL and the LM regimes are too
close for the sign-change to occur, and the crossover is indicated by
a shallow minimum between two asymmetric peaks or just a shoulder in
$S(T)$. For larger $\Gamma$-values, the crossover is from the FL to a
MV regime but the FL scale is large, $T_0\propto g$, and $S(T)$ grows
monotonically towards the high-temperature maximum. The variation of
the hybridisation at constant temperature changes the shape of the
spectral function from a typical LM to a typical FL shape. 

The phase diagram deduced from the single-impurity Anderson model
calculations is in rather good agreement with the experimental one. In
the inset to Fig.~\ref{fig:theo_phasediagram} the pressure dependence
of the experimentally obtained $T_K$-values of CeRu$_2$Ge$_2$ (already
shown in Fig.~\ref{fig:phasediagram}) are re-plotted together with the
data of CeRu$_2$Si$_2$. The $T_K(p)$-data of CeRu$_2$Si$_2$ were
calculated from the measured $A(p)$-dependence
\cite{Payer93,Mignot89,Thompson85} using the scaling relation
$T_K\propto \sqrt A$ with the assumption $T_K=24$~K at ambient
pressure \cite{Besnus85}.  In order to compare the CeRu$_2$Si$_2$ data
with those of CeRu$_2$Ge$_2$, the former had to be shifted by 8.4~GPa
\cite{Wilhelm04}. On the verge of the magnetic instability ($p\simeq
7.8$~GPa) but still in the magnetic phase, the $T_K(p)$-variation
detected with the $S(T)$ experiment, seems to confirm the exponential
increase of $T_0$ with $\Gamma$. The change of the
$T_0(\Gamma)$-dependence at larger $\Gamma$ is also seen in the
experimental data above the magnetic instability, where the FL regime
is entered. Here, the $T_K(p)\propto \sqrt{A(p)}$ variation for
CeRu$_2$Ge$_2$ as well as CeRu$_2$Si$_2$ is different to the $T_K(p)$
dependence deduced from the $S(T)$-data of CeRu$_2$Ge$_2$
(Fig.~\ref{fig:phasediagram} and inset to
Fig.~\ref{fig:theo_phasediagram}). The crossover into the MV regime,
where $T_0\propto \tilde{E}_f$, yields to a reduced pressure
dependence (data only for CeRu$_2$Si$_2$) in agreement with the
calculated $T_0(\Gamma)$-dependence for $\Gamma>170$~meV. It is
interesting to note that the characteristic energy scales of a local
and a coherent FL do not seem to differ very much, so that the
universal power laws of a coherent FL provide an estimate of the
single-impurity scale ${T_0}$. However, a proper treatment of the
low-temperature and high-pressure properties requires a lattice model.

The NCA used to describe the pressure-induced features in
CeRu$_2$Ge$_2$ is not restricted to this compound. Other ternary
Ce-compounds like CeCu$_2$Si$_2$ or CePd$_2$Si$_2$ can also be
described, if the parameters are adjusted appropriately. The method
works also for Yb-compounds if the {\em decrease} of the coupling
between the 4f-states and the conduction band with pressure is taken
into account.

\section{Conclusions}
A qualitative understanding of the pressure-induced changes in the
electrical resistivity $\rho(T) $ and thermopower $S(T)$ of
CeRu$_2$Ge$_2$ was developed in the framework of the single-site
Anderson model. The evolution of $\rho(T) $ and $S(T)$ with pressure
was accounted for by the increase of the 4f-conduction band electron
hybridisation $\Gamma$. The calculated spectral functions $A(\omega)$
show that the temperature-induced redistribution of spectral weight
yields several of the pronounced features observed in the measured
transport quantities. The position of the peak in $A(\omega)$ close to
the chemical potential gives a very reliable estimate of the
characteristic Kondo temperature $T_0$ at low temperature. The
crossovers between various fixed points of the Anderson model and the
redistribution of the single-particle spectral weight within the Fermi
window explain the hybridisation dependence of the transport
coefficients and the seemingly complicated temperature and pressure
dependences of experimental $S(T)$-data. The $\Gamma$-dependence of
the characteristic temperature scale $T_0$ is in qualitative agreement
with the experimentally determined $T_K(p)$-variation.

\begin{acknowledgements}
We thank T.~C.~Kobayashi and M.~Malquarti for assistance during the thermopower
measurements. H. W. is grateful to B. Schmidt for many stimulating
discussions. The work was partly supported by the Swiss National Science
Foundation.
\end{acknowledgements}

\end{document}